\documentstyle[epsfig,longtable,color]{aipproc}

\definecolor{purp}{cmyk}{.2,1,0,.4}
\definecolor{grnn}{cmyk}{1,.2,.5,.4}
\definecolor{brown}{rgb}{.5,.35,.1}

\newcommand{\lae}{\stackrel{<}{\sim}}
\newcommand{\gae}{\stackrel{>}{\sim}}

\begin{document}
\null\vspace{-1cm}\hfill{\bf BUHEP-97-22}\break

\title{The Significance of the \\ Heavy Top Quark\footnote{Talk given at
    Beyond The Standard Model V, 29 April -- 4 May, 1997,
    Balholm, Norway.}}

\author{Elizabeth H. Simmons}
\address{Department of Physics, Boston University\\
590 Commonwealth Avenue\\
Boston, MA  02215}
%\lefthead{LEFT head}
%\righthead{RIGHT head}
\maketitle

\null\vspace{-.5cm}
\begin{abstract}
Experiment shows that the top quark is far heavier than the other elementary
fermions.  This finding has stimulated research on theories of
electroweak and flavor symmetry breaking that include physics beyond
the standard model. Efforts to accommodate a heavy top quark within
existing frameworks have revealed constraints on
model-building. Other investigations have started from the premise
that a large top quark mass could signal a qualitative difference
between the top quark and other fermions, perhaps in the form of new
interactions peculiar to the top quark. Such new dynamics may also
help answer existing questions about electroweak and flavor
physics. This talk explores the implications of the heavy
top quark in the context of weakly-coupled (e.g. SUSY) and
strongly-coupled (e.g. technicolor) theories of electroweak symmetry
breaking.

\end{abstract}

\null\vspace{-.5cm}
\section*{Introduction}

Two outstanding mysteries in particle theory are the cause of
electroweak symmetry breaking and the origin of flavor symmetry breaking
by which the quarks and leptons obtain their diverse masses.  The
Standard Model of particle physics, based on the gauge group $SU(3)_c
\times SU(2)_W \times U(1)_Y$ accommodates both symmetry breakings by
including a fundamental weak doublet of scalar (``Higgs'') bosons ${\phi
  = {\phi^+ \choose \phi^0}}$ with potential function $V(\phi) = \lambda
\left({\phi^\dagger \phi - \frac12 v^2}\right)^2$.  However the Standard
  Model provides no explanation of the dynamics responsible for the
  generation of mass.  Furthermore, the scalar sector suffers from two
  serious problems.  The scalar mass is unnaturally sensitive to the
  presence of physics at any higher scale $\Lambda$ (e.g. the Planck
  scale or a grand-unification scale):
\begin{center}
\begin{math}
 \hspace{0cm}{\hbox{\epsfysize=.4in 
    \epsfbox{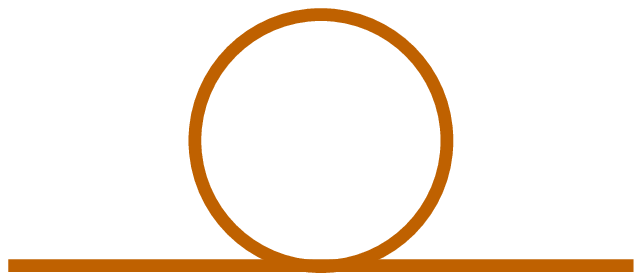}}} \hspace{1cm}
\raise12pt\hbox{$ \Rightarrow
 \ M_H^2\ \propto\ \Lambda^2$}
\end{math}
\end{center}
This is known as the gauge hierarchy problem.  In addition, if the
scalar must provide a good description of physics up to arbitrarily high
scale (i.e., be fundamental, not composite), the scalar's self-coupling
($\lambda$) is driven to zero
\begin{center}
\begin{math}
 \hspace{0cm}{\hbox{\epsfysize=.4 in 
    \epsfbox{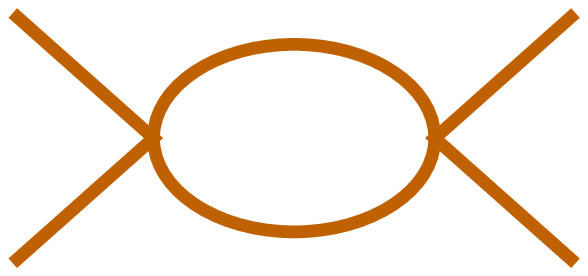}}} \hspace{1cm}
\raise12pt\hbox{$\Rightarrow\
  \beta\ =\ {{3\lambda^2}\over{2\pi^2}}\ > \ 0$}
\end{math}
\end{center}
at finite energy scales.  That is, 
the scalar field theory is free (or ``trivial'').  Then the scalar 
cannot fill its intended role:
 if $\lambda = 0$, the electroweak symmetry is not spontaneously
broken.  It is, thus, necessary to seek the origin of mass in physics
that lies beyond the Standard Model.

One intriguing possibility is to introduce
supersymmetry \cite{ehs-1-susy}.  The gauge structure of the minimal 
supersymmetric
version of the Standard Model (MSSM) is identical to that of the
Standard Model, but each ordinary fermion (boson) is paired
with a new boson (fermion) called its ``superpartner'' and two Higgs
doublets are needed to provide mass to all the ordinary fermions.
Each loop of ordinary particles contributing to the higgs boson's
mass is now countered by a loop of superpartners. If the masses 
of the ordinary 
\begin{center}
\begin{math}
 \hspace{3cm}{\hbox{\epsfysize=.5 in 
    \epsfbox{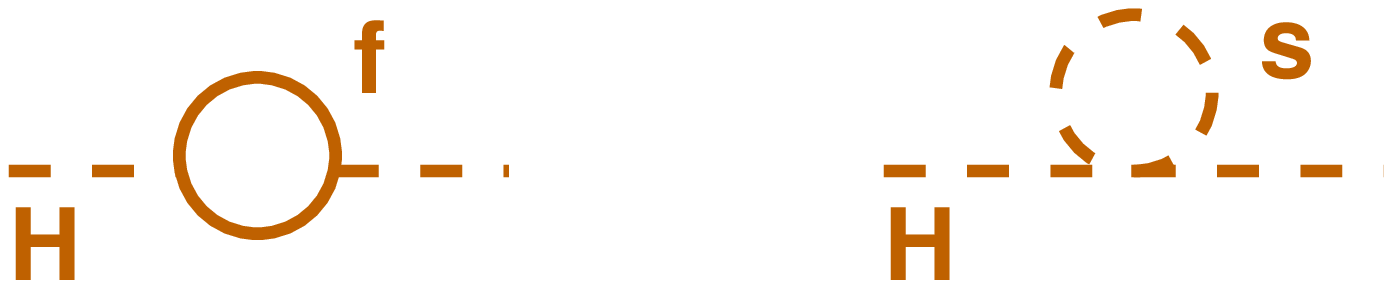}}} \hspace{1cm}
\raise15pt\hbox{$\Rightarrow \ \delta M_H^2\ \sim\ 
\frac{g_f^2}{4\pi^2} (m_f^2 - m_{s}^2) $}
\end{math}
\end{center}
\noindent particles and superpartners
are sufficiently close, the gauge hierarchy can be stabilized
\cite{ehs-1-stab} .  In addition, supersymmetry relates the scalar
self-coupling to gauge couplings, so that triviality is not a concern.

Another interesting class of models involve dynamical electroweak
symmetry breaking\cite{ehs-1-dynam}.  In these theories, a new strong gauge interaction
with $\beta < 0$ (e.g technicolor) breaks the chiral symmetries of a
set of massless fermions $f$ at a scale $\Lambda \sim 1$TeV.  If the
fermions carry appropriate electroweak quantum numbers, the resulting
condensate $\langle \bar f_L f_R \rangle \neq 0$ breaks the
electroweak symmetry as desired.  The logarithmic running of the
strong gauge coupling renders the low value of the electroweak scale
(i.e. the gauge hierarchy) natural.  The absence of fundamental scalar
bosons obviates concerns about triviality.

How is one to choose among the various models ?  Consider a rough
graph of the masses of the known fermions and gauge bosons:
\vspace{.9cm}
{\color{green}\bf \hrule}
\vspace{-1cm}

\begin{tabbing}
\hskip1cm \=\hskip 2cm \= \hskip 2cm \= \hskip 2cm \= \hskip 2cm \=
\hskip 2cm \= \kill
\hskip .5cm{\color{blue}e}\>{\color{blue}\ \ u \ \ \ \ \ \ d} \>
\>{\color{purp} $\mu$\hspace{.133cm}{
   s}} \> {\color{purp} c} {\color{red} ${\tau}$\ \ \ \ \ \ b} 
\>\hskip 1.27cm {\color{black} W Z} \>\hskip .67cm{\color{red} t}\\
\>{\color{green}$\vert$} \>{\color{green}$\vert$} \>{\color{green}$\vert$}\> 
{\color{green}$\vert$} \>{\color{green}$\vert$} \> {\color{green}$\vert$}\\
\>{\color{green} .001 }\>{\color{green} .01 }\>{\color{green} .1 }
\>{\color{green} 1} \>{\color{green} 10}\>{\color{green} 100 GeV }\\
\end{tabbing}
 \vspace{-.5cm}
 The top quark is singled out\cite{ehs-1-top}: it is the heaviest known
 elementary particle, with a mass of order the electroweak scale, $v
 \sim 246$GeV, and is far heavier than its weak partner ($b$).  This
 suggests that the top quark may afford us insight about existing models
 of electroweak physics and may even play a special role in electroweak
 and flavor symmetry breaking.

\section*{The Top Quark Makes A Difference}

The large mass of the top quark has illuminated aspects of existing
theories of electroweak and flavor physics.  We review some
opportunities and constraints the top quark has provided for
supersymmetric models and theories of dynamical electroweak symmetry
breaking.

\subsection*{Electroweak Symmetry Breaking in SUSY Models}

A challenge for supersymmetric models is to explain why the Higgs scalar
develops a negative mass-squared (so that the electroweak symmetry
breaks) while the scalar partners of the ordinary
fermions do not (so that color and electromagnetism are preserved).  The
heavy top quark provides a solution.  

In many types of models (e.g. the constrained MSSM (CMSSM), models of
dynamical supersymmetry breaking) the mass-squared of the higgs
is related to that of the sfermions, and is therefore positive at scales
($M_X$) well above the weak scale\cite{ehs-2-susyy}  
\begin{displaymath}
M_{h,H}^2(M_X) = m_0^2 + \mu^2 \ \ \ \ \ \ \ \ 
M_{\tilde f}^2(M_X) = m_0^2
\end{displaymath}
\noindent Moving towards lower scales, the masses of the higgs ($M_h$) 
and of the top squarks ($\tilde{M_{t_R}}, \tilde{M_{Q^3_L}}$) evolve 
under the renormalization group\cite{ehs-2-evolve}:
\begin{displaymath}
{\displaystyle d\over{\displaystyle d \ln(q)}} 
\pmatrix{{\textstyle M_h^2} \cr {\textstyle\tilde{M}_{t_R}^2}\cr
{\textstyle\tilde{M}_{Q^3_L}}^2\cr} = - {\displaystyle
8\alpha_s\over{\displaystyle 3\pi}} M_3^2
\pmatrix{\textstyle 0 \cr 1 \cr 1 \cr}
{{\lambda_t^2}\over{\displaystyle 8\pi^2}}\, 
({\tilde{M}_{Q^3_L}}^2 +
{\tilde{M}_{t_R}}^2 + M_h^2 + A_{o,t}^2) \textstyle \pmatrix{{\textstyle
    3} \cr \textstyle 2\cr \textstyle 1\cr}
\end{displaymath}
Clearly, the top quark's large Yukawa coupling
$\lambda_t$ is important and the mass-squared of the higgs $h$ 
is affected more than that of the squarks.  The approximate solution for
the light higgs mass-squared at scale $q$ is
\begin{displaymath}
M_h^2 (q) = M_h^2 (M_X)
 - {3\over{8\pi^2}} 
{\lambda_t^2}  \left( {\tilde{M}_{Q^3_L}}^2 +
{\tilde{M}_{t_R}}^2 + M_h^2 + A_{o,t}^2\right) ln \left({M_X \over q } 
\right)
\end{displaymath}
For a top quark mass $m_t \sim 175$ GeV, the higgs mass-squared is
driven negative near the electroweak scale \cite{ehs-2-evolve}, while
those of the squarks are not.  As desired, the electroweak symmetry
breaks while color and electromagnetism survive.

\subsection*{Effects on the Higgs Spectrum of SUSY Models}

After electroweak symmetry breaking, the higgs spectrum of the MSSM
includes two neutral scalar bosons.  The tree-level upper bound on the
mass of the lighter one ($h$) is $M_h < M_Z \vert\cos(2\beta)\vert$.
This would appear to forbid $\tan\beta \sim 1$ and lies quite close
to the experimental lower bound on $m_h$.

Radiative corrections involving the heavy top quark and its
superpartners provide a resolution.  For $\tan\beta >
1$, the bound on $M_h$ becomes \cite{ehs-2-higgsm}
\begin{displaymath}
M_h^2 \lae M_Z^2 \cos^2(2\beta) + {3 G_f
    \over{\sqrt{2}\pi^2}}\ {m_t^4}\ ln
\left({\tilde m^2 \over m_t^2}\right)
\end{displaymath}
so that $M_h \lae 130$ GeV and $\tan\beta \sim 1$ (the $U(1)_R$ symmetric
limit \cite{ehs-2-mr} or ``light gaugino-higgsino window''
\cite{ehs-2-lgh}) is still viable.  

\subsection*{Oblique Corrections in Dynamical Models}

Extended technicolor (ETC) is an explicit realization of dynamical
electroweak symmetry breaking and fermion mass generation. One starts
with a strong gauge group (technicolor) felt only by a set of new
massless fermions (technifermions) and extends the technicolor gauge
group to a larger (ETC) group under which ordinary fermions are also
charged.  At a scale $M$, ETC breaks to its technicolor subgroup and the
gauge bosons coupling ordinary fermions to technifermions acquire a mass
of order $M$.  At a scale $\Lambda_{TC} < M$ the technicolor coupling
becomes strong enough to form a technifermion condensate and break the
electroweak symmetry.  Because the massive ETC gauge bosons couple the
ordinary fermions to the condensate, the ordinary fermions acquire mass
too.  The top quark's mass, e.g.,  comes from 
\null\begin{center}
\begin{math}
 {\hbox{\epsfysize=.7 in 
    \epsfbox{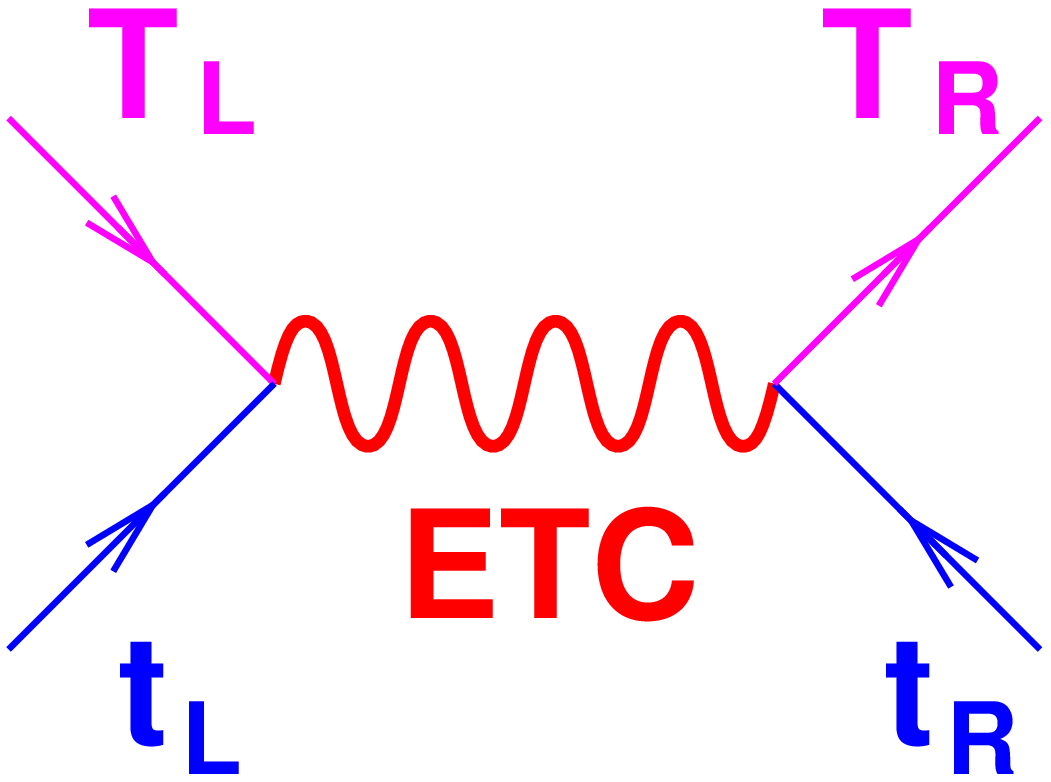}} } \hspace{2cm}
\raise20pt\hbox{\LARGE ${\bf\Longrightarrow}$} 
\hspace{1cm} 
 \ {\hbox{\epsfysize .7 in
    \epsfbox{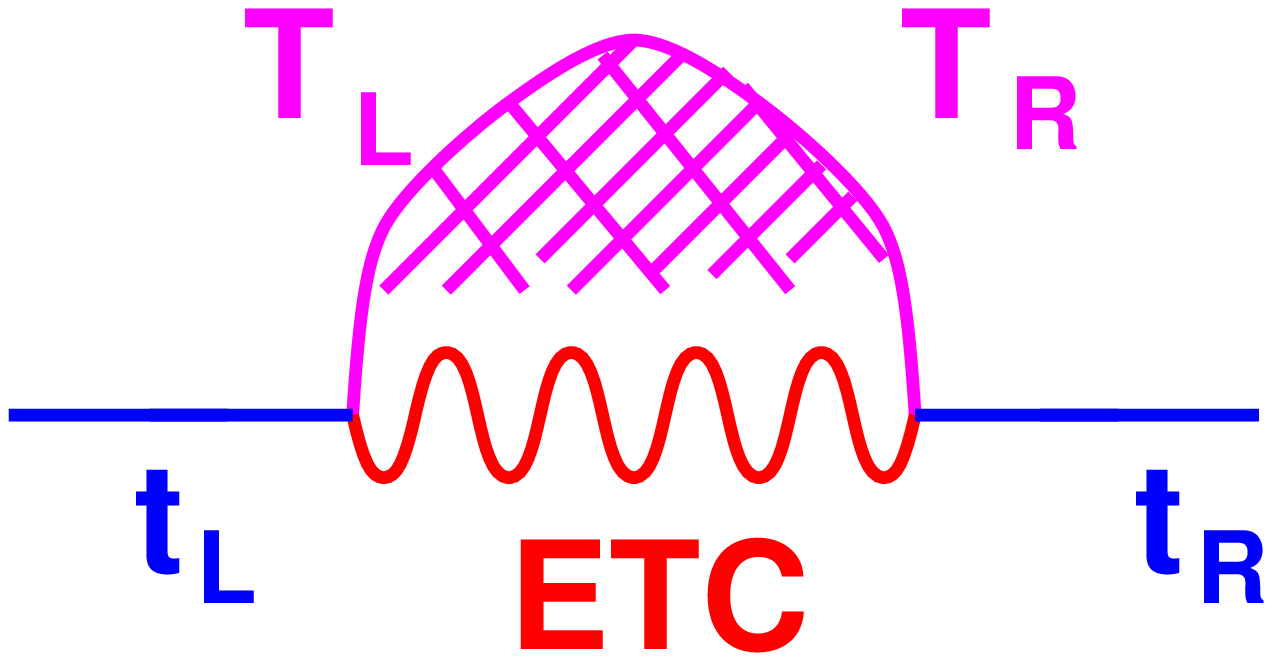}} }  
\end{math} \end{center} 
\noindent and its size is $m_t \approx$
$(g_{ETC}^2/M^2)\langle T\bar T\rangle \approx (g_{ETC}^2/M^2)(4\pi v^3)$.
Thus the scale $M$ must satisfy $M/g_{ETC} \approx 1.4$ TeV in order to
produce $m_t = 175$ GeV.

Several difficulties arise when one tries to balance the need to create
a wide range of ordinary fermion masses against the requirement of
keeping the oblique correction $\Delta\rho$ small.  First are the
so-called ``direct'' contributions\cite{ehs-2-dirr} to $\Delta\rho$. The
ETC sector must violate weak isospin in order to make $m_t \gg m_b$.
This can induce dangerous technifermion ($\Psi$) contributions to
$\Delta\rho$:
\vskip .5cm
\begin{math}
\hspace{-.3cm}{\hbox{\epsfysize=.5truein
    \epsffile{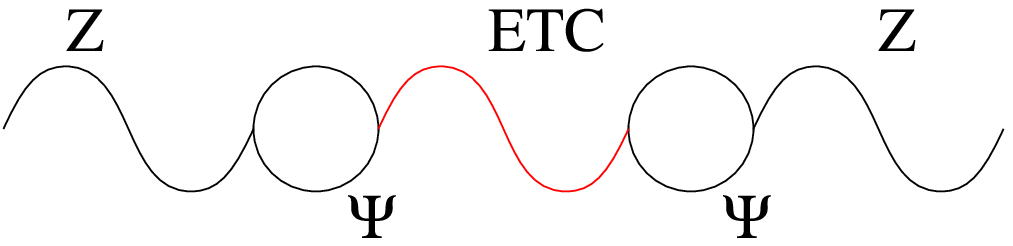}}} \raise20pt\hbox{$
\hspace{.8cm} \Rightarrow \hspace{.8cm} 
\Delta\rho \approx  12\% \cdot 
\left({\sqrt{N_D} F_{TC} \over 250 {\rm\ GeV}}\right)^2
\cdot \left({1 {\rm\ TeV} \over M/g_{ETC}}\right)^2.$}
\end{math}
\noindent To satisfy the experimental constraint 
$\Delta\rho \leq 0.4\%$, one might consider making the ETC boson heavy.
This requires $(M/g_{ETC}) > 5.5 {\rm TeV} (\sqrt{N_D}F_{TC}/250{\rm
  GeV})$, which is too heavy to produce $m_t \sim 175$GeV.  A better
alternative\cite{ehs-1-dynam} is to arrange for separate dynamical
 sectors to break the
electroweak symmetry and produce the bulk of $m_t$.  Then the
technifermions contributing to $\Delta\rho$ can satisfy
$\sqrt{N_D}F_{TC} \ll 250$GeV.

\begin{figure}[h] % fig 1
\centerline{\epsfig{file=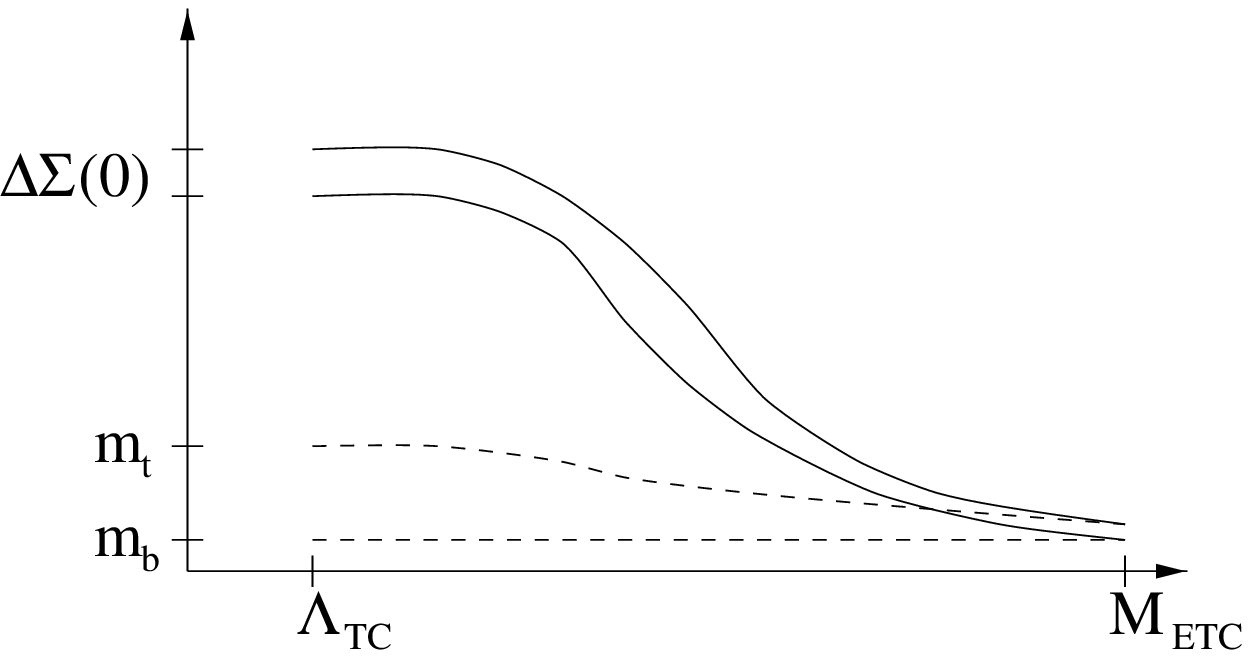,height=1.4in,width=3.in}}
\vspace{10pt}
\caption{Dynamical technifermion and fermion masses sketched as a
  function of momentum.  This illustrates [3] the
  scenario where the bulk of $m_t$ and $m_b$ comes from a new strong
  interaction not felt by the technifermions or other quarks.}
\label{figyy}
\end{figure}

There are also ``indirect'' contributions\cite{ehs-1-dynam} to
$\Delta\rho$ from isospin- violating 
\null\vskip.1cm
\begin{math}
\hspace{-\parindent}\hbox{\parbox{2.75in}{\rm splittings between technifermion 
dynamical masses $\Sigma_U(0), \Sigma_D(0)$.   
Again, a solution is to have the $t$ and $b$ quarks obtain 
only part of
their masses from technicolor.  This can keep the technifermion mass
splitting small enough }}
\hspace{1.5cm} \lower25pt\hbox{\epsfysize=.7 in 
    \epsfbox{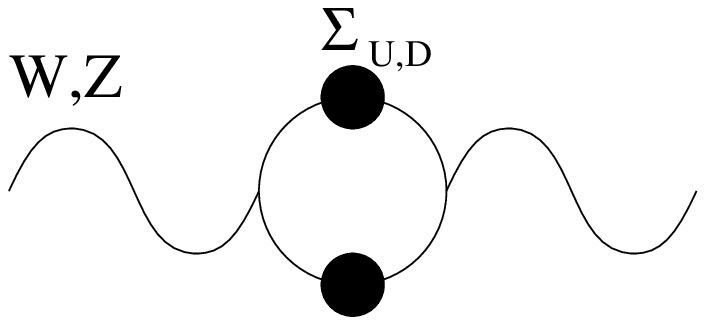}} \\  
\end{math}
\noindent at low energies $\Delta \Sigma(0) \simeq
m_t(M_{ETC})-m_b(M_{ETC}) \ll m_t(0) - m_b(0)$.  However the $t$ and $b$
quarks must feel some additional strong interaction not shared by the
light fermions or technifermions, which can generate $m_t \gg m_b \gg
m_{light\ quark}$ as sketched in Figure \ref{figyy}.   
We will revisit these ideas later.

\subsection*{Corrections to $Zb\bar b$ in Dynamical Models} 

%
%In discussing oblique corrections, we related the isospin violation
%exhibited by the ETC boson that generates $m_t$ to isospin violation
%elsewhere in the ETC sector.  The contributions to $\Delta\rho$ are
%physically distinct from the generation of $m_t$ and arise due to other
%ETC bosons -- and it may therefore be possible to build a model that
%circumvents the related constraints. However, the ETC boson that 
%generates $m_t$ also has a direct effect on another precisely-measured
%observable: $R_b$.
%
While the contributions to $\Delta\rho$ just discussed are physically
distinct from generation of $m_t$, the very ETC boson responsible
for $m_t$ makes potentially large contributions to $R_b$.  Consider the
simplest ETC models, those in which the ETC and
weak gauge groups commute and the ETC
bosons carry no weak charge.  The ETC boson responsible for generating
the top quark mass couples to the current $\xi (\bar T_L \gamma_\mu
\psi_L) + \xi^{-1} (\bar U_R \gamma_\mu t_R)$ where $\xi$ is an ETC
Clebsch, $\psi \equiv (t,b)$, and $T \equiv (U, D)$ is a technifermion
doublet.  The top quark mass comes from
\begin{center}
\begin{math}
 {\hbox{\epsfysize=.7 in 
    \epsfbox{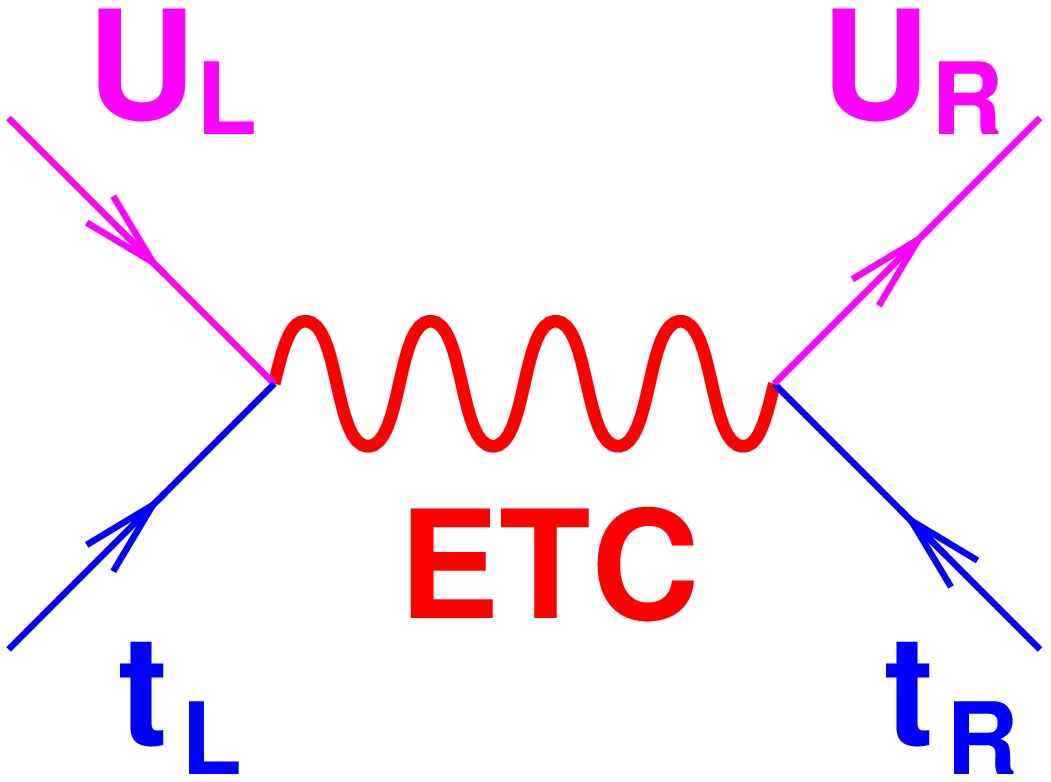}} } \hspace{2cm}
\raise20pt\hbox{\LARGE ${\bf\Longrightarrow}$} 
\hspace{1cm} 
 \ {\hbox{\epsfysize .7in
    \epsfbox{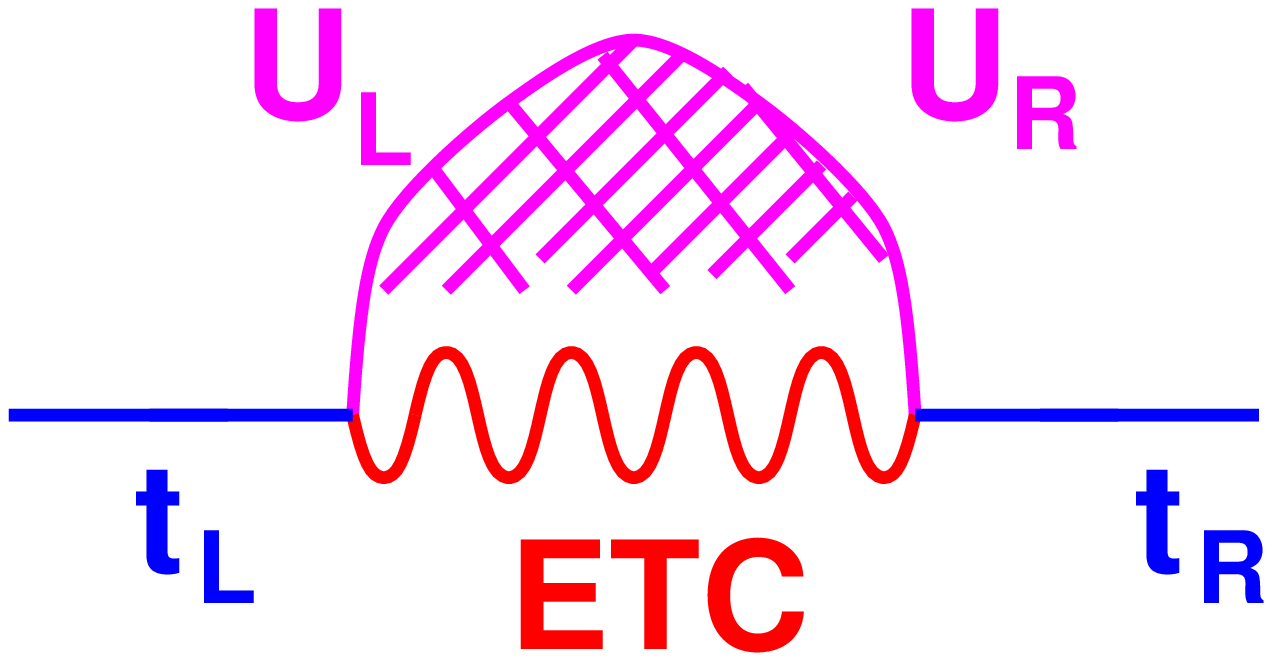}} }  
\end{math}

\end{center}
and is of order $m_t \equiv (g^2_{ETC}/M^2)(4\pi v^3)$.  This implies
\cite{ehs-2-sbs} 
that the size of the typical vertex correction arising from exchange of
this ETC boson is proportional to the top mass: ${g_{ETC}^2 v^2 / {M^2}}
\approx {m_t / {4 {\pi} v}}$.

In particular, this ETC boson causes a radiative correction to the $Zb\bar b$
vertex
\null\vskip.1cm
\begin{math}
\hspace{-\parindent}\hbox{\parbox{2.75in}{\rm that shifts 
   the $Zb\bar b$ coupling by
            $\delta g_L = {e \over {4
    \sin{\theta} \cos{\theta}}} {g_{ETC}^2 v^2 \over {M^2}} $ .  As we
    have seen, this is proportional
    to $m_t$.  Because $m_t$ is so large,
    $R_b \equiv \Gamma(Z \to b\bar b) / \Gamma(Z \to {\rm hadrons})$ 
    is decreased by $\approx 5\%$ relative to the standard model
    prediction; such a value of $R_b$ has been excluded by
    experiment\cite{rbvalue}.  Hence, these
``commuting'' ETC models are ruled out\cite{ehs-2-sbs}.}}
\hspace{1.5cm} \lower40pt\hbox{\epsfysize=1.25 in 
    \epsfbox{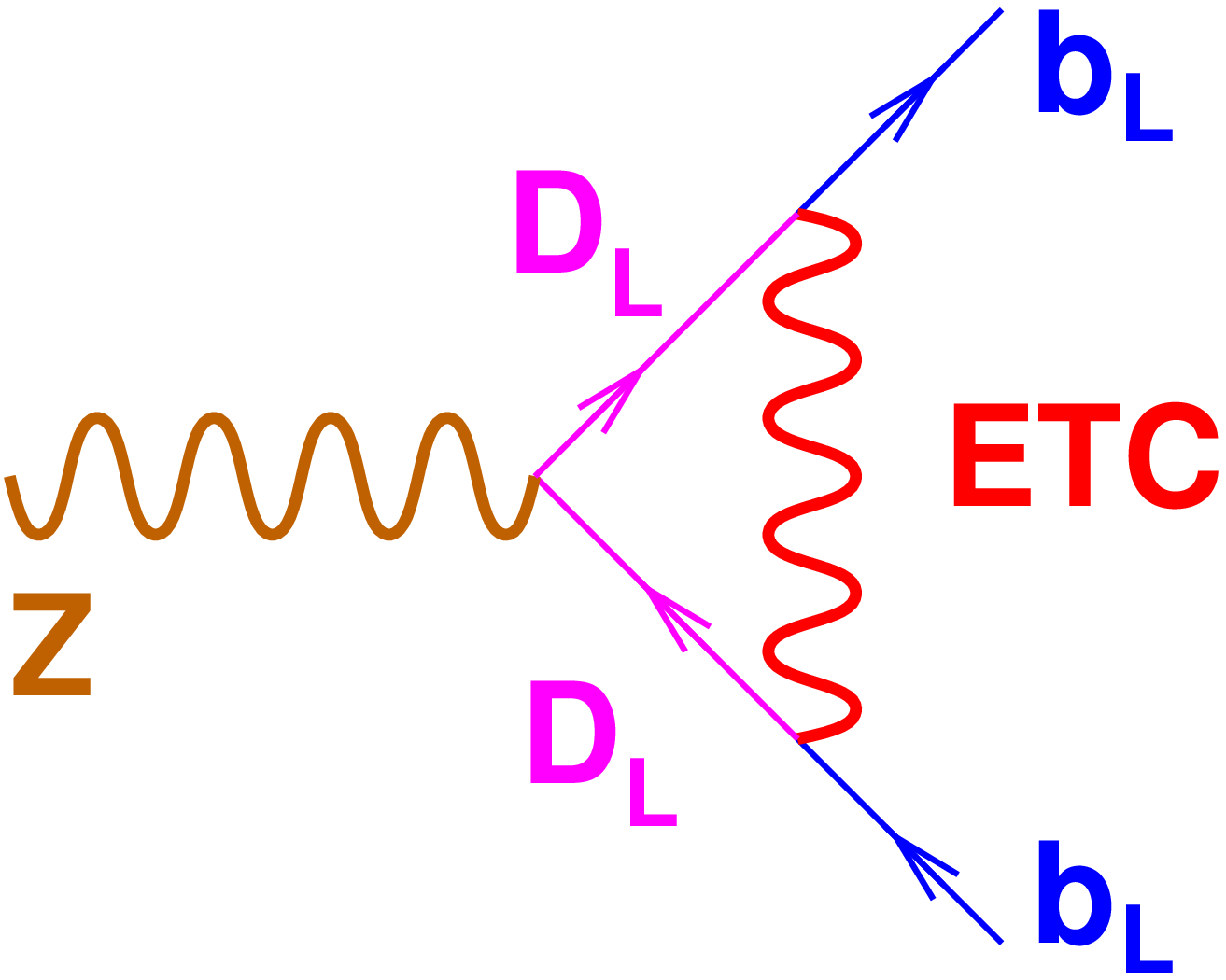}} \\  
\end{math}

\section*{The Top Quark May Be Different}

We next consider whether the top quark's large mass implies that the $t$
quark has unique new interactions.   Such scenarios 
provide alternative mechanisms of electroweak and flavor symmetry
breaking and are experimentally testable.

\subsection*{New Weak Interactions for the Top Quark}

To start, we return to extended technicolor.  Our discussion of $R_b$
did not rule out the possibility of ``non-commuting'' models in which
$SU(2)_W$ is embedded in $G_{ETC}$ so that the ETC bosons carry weak
charge.  To build such a model, one must balance the requirements of
providing a range of quark masses against the need to ensure that the
weak interactions are universal at low energies.  As discussed in
\cite{ehs-2-ncetc}, this leads to separate weak interactions for the 3rd
generation fermions ($SU(2)_{heavy}$) and the light fermions
($SU(2)_{light}$) and the symmetry-breaking pattern
\begin{eqnarray*}
G_{ETC}  &\times& SU(2)_{light}\\
&\downarrow& \\
G_{TC} \times SU(&\!\!\!2&\!\!\!)_{heavy}  \times SU(2)_{light} \\
&\downarrow&\\
G_{TC}  &\times& SU(2)_{weak}
\end{eqnarray*}
The result is a model where the top quark has non-standard weak
interactions. 

Our first concern is the value of $R_b$ this model predicts.
The ETC boson 
\null\vskip.1cm
\begin{math}
\hspace{-\parindent}\hbox{\parbox{2.75in}{\rm responsible for 
creating $m_t$ is now a weak
doublet coupling to $\xi (\bar U_L \gamma_\mu
\psi_L) + \xi^{-1} (\bar T_R \gamma_\mu t_R)$.  This gives a direct
radiative correction to the $Zb\bar b$ vertex which is of the same
magnitude but opposite sign to the correction in commuting models
\cite{ehs-2-ncetc}.  That alone would }}
\hspace{1.5cm} \lower40pt\hbox{\epsfysize=1.25 in 
    \epsfbox{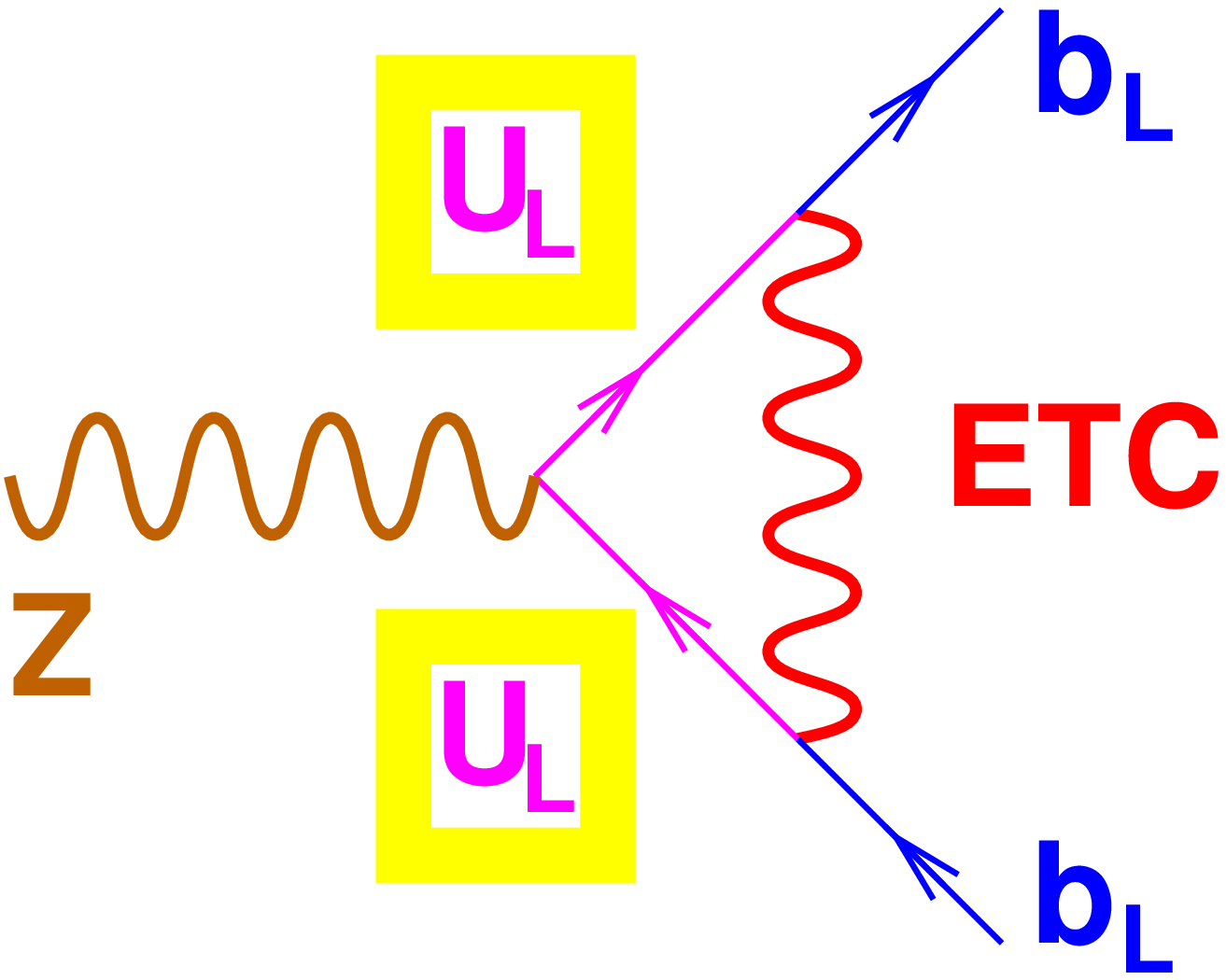}} \\  
\end{math}
\noindent not make $R_b$ compatible with experiment.  But in
addition, mixing of the $Z$
\null\vskip.1cm
\begin{math}
  \hspace{-\parindent}\hbox{\parbox{2.75in}{\rm bosons from the
      two weak gauge groups gives another correction to the $Zb\bar b$
      vertex which essentially brings $R_b$ back to the standard model
      value.  Thus non-commuting ETC can be consistent with the measured
      value of $R_b$.}} 
\hspace{1.5cm} \lower25pt\hbox{\epsfysize=.8 in
    \epsfbox{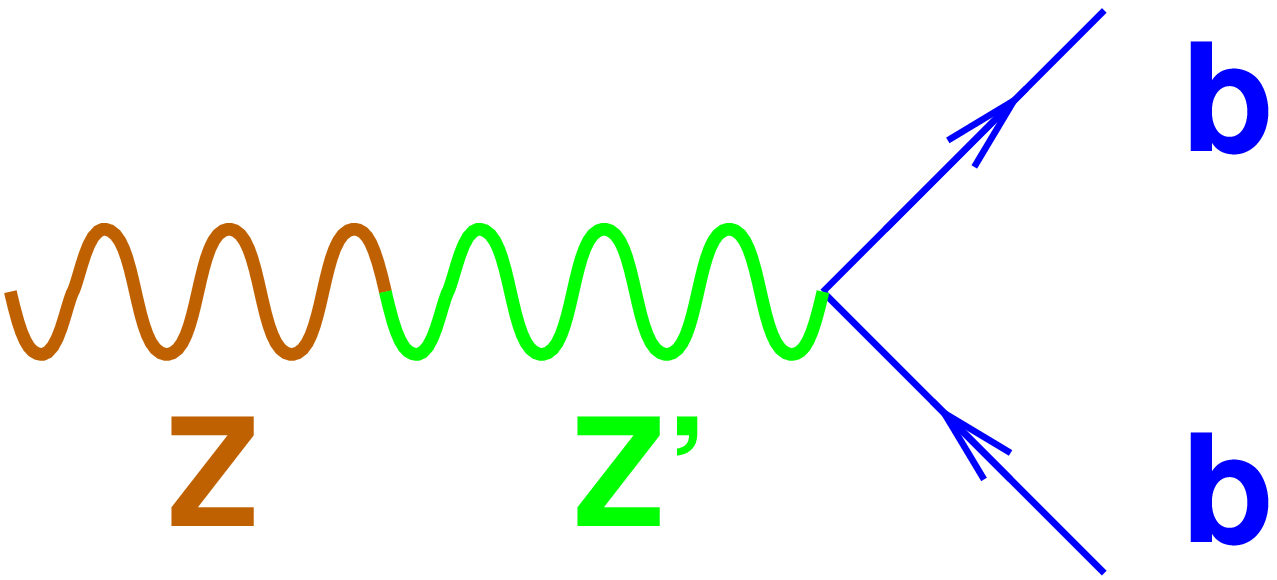}} \\ 
\end{math}

\begin{figure}[h] % fig 2
\centerline{\epsfig{file=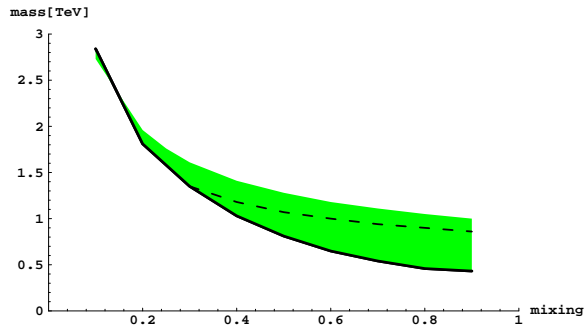,height=1.65in,width=3.in}}
\vspace{10pt}
\caption{The axes are heavy $W$ mass and degree of mixing of the two weak gauge
  groups in a non-commuting ETC model.  The shaded region is the area
  consistent with low-energy data in which $R_\sigma$ would be increased
  by at least 16\% [13].}
\label{figXX}
\end{figure}

Non-standard top quark weak interactions may be detectable in
single top-quark production at TeV33 \cite{ehs-3-wtb}.  The ratio of
cross-sections $R_\sigma \equiv \sigma(\bar{p}p \to t b) /
\sigma(\bar{p}p \to l \nu)$ can be measured (and calculated) to an
accuracy \cite{ehs-3-wtbacc} of at least $\pm 8\%$.  A non-commuting ETC model
might alter $R_\sigma$ in several ways.  Mixing of the two $W$ bosons
alters the light $W$'s couplings to the final state fermions.
Exchange of both heavy and light $W$ bosons contributes to the
cross-sections.  But exchange of the ETC boson that generates $m_t$
does {\bf not} modify the $Wtb$ vertex, because the boson does not
couple to all of the required fermions: ($t_R, b_R, U_R, D_R$).  In the
case of $R_b$, the vertex and boson mixing effects canceled, leaving
$R_b$ at the standard model value; here the boson mixing effects are
not canceled and can yield a visible {\bf increase} in $R_\sigma$
(see Figure \ref{figXX}). 
\null\vspace{-.5cm}

\subsection*{New Strong Interactions for the Top Quark}

To build a dynamical symmetry breaking model that provides both
electroweak symmetry breaking and a large $m_t$, it has been suggested
\cite{ehs-3-ttcond} 
that all or some of electroweak symmetry breaking could be caused by a
top quark condensate ($ \langle \bar{t}t\rangle \neq 0$).  One way to
implement this is to start with a spontaneously broken strong gauge
interaction that distinguishes top from the other quarks.  
Suppose the model includes an $SU(3)_H$ for the $t$ (and $b$) and an
$SU(3)_L$ for the other quarks which break to their diagonal
subgroup (identified with $SU(3)_{QCD}$ at a scale $M$.  At
energies below $M$,  exchange of the heavy
gauge bosons yields a new four-fermion interaction  
can cause top quark condensation.
\begin{displaymath}
{\cal L} \ \supset\  - {4\pi\kappa \over M^2}
\left(\overline{t}\gamma_\mu {\lambda^a\over 2} t\right)^2
\end{displaymath}

The simplest ``topcolor-assisted technicolor'' model\cite{ehs-3-tc2}
incorporating top condensates has the following gauge group and
symmetry-breaking pattern.
\begin{eqnarray*}
G_{TC} \times {SU(3)_H \times SU(3)_L} &\times& SU(2)_W
  \times  {U(1)_H \times U(1)_L} \\
&\downarrow& \ \ M \gae 1 TeV \\
G_{TC} \times {SU(3)_C}  &\times& SU(2)_{W} \times
  {U(1)_Y} \\
&\downarrow&\ \ \ \Lambda_{TC}\sim 1 TeV \\
{SU(3)_C} &\times& {U(1)_{EM}}
\end{eqnarray*}
The groups $G_{TC}$ and $SU(2)_W$ are
ordinary technicolor and weak interactions; the strong and hypercharge
groups labeled ``H'' couple to 3rd-generation fermions and have stronger
couplings than the ``L'' groups coupling to light fermions
The separate $U(1)$ groups ensure that the bottom quark will not
condense when the top quark does.  Below the scale $M$, the Lagrangian
includes effective interactions for $t$ and $b$:
\begin{displaymath} 
-{{4\pi \kappa_{tc}}\over{M^2}} \left[\overline{\psi}\gamma_\mu  
{{\lambda^a}\over{2}} \psi \right]^2 
-{{4\pi \kappa_1}\over{M^2}} \left[{1\over3}\overline{\psi_L}\gamma_\mu  
\psi_L + {4\over3}\overline{t_R}\gamma_\mu  t_R
-{2\over3}\overline{b_R}\gamma_\mu  b_R
\right]^2
\end{displaymath}
So long as the following relationship is satisfied (where the critical
value is $\kappa_c \approx 3\pi/8$ in the NJL approximation \cite{ehs-3-njl}) 
\begin{displaymath}
\kappa^t = \kappa_{tc} +{1\over3}\kappa_1 >
\kappa_c  >
\kappa_{tc} -{1\over 6}\kappa_1 = \kappa^b
\end{displaymath}
only the top quark will condense and become very massive\cite{ehs-3-tc2}.

Topcolor-assisted technicolor
models  have
several appealing
features\cite{ehs-3-tc2phen}\cite{ehs-1-dynam}. Technicolor 
causes most of the electroweak
symmetry breaking, with the top condensate contributing a decay constant
$f \sim 60$ GeV; this prevents $\Delta\rho$ from being too large, as
mentioned earlier.  So long as the $U(1)_H$ charges of the
technifermions are isospin-symmetric, they cause no additional large
contributions to $\Delta\rho$.  ETC dynamics at a scale $M \gg 1$TeV
generates the light fermion masses and contributes about a GeV to the
heavy fermions' masses; this does not generate large corrections to
$R_b$.  The top condensate can, then, provide the bulk of the top quark
mass.  Precision electroweak data constrain the mass of the extra $Z$
boson in these models to weigh at least 1-2 TeV \cite{ehs-3-rscjtbd}.

\subsection*{Light Top Squarks}

Finally, we return to supersymmetric models.  Consider the mass matrix
for the supersymmetric partners of the top quark:
\begin{center}
\begin{math}\null\hspace{-1cm}\tilde{m}_t ^2 = 
\pmatrix{\tilde{M}^2_Q + m_t^2 &\ & 
{m_t}(A_t + \mu\cot\beta)\cr 
+ M_Z^2(\frac12 - \frac23 \sin^2\theta_W)\cos2\beta &\ &\cr
\ &\ &\ \cr  {m_t}(A_t +
    \mu\cot\beta)& &\tilde{M}^2_U + m_t^2 \cr
&\ & + \frac23 M_Z^2 \sin^2\theta_W
    \cos2\beta\cr} 
\end{math}
\end{center}
the presence of $m_t$ in the off-diagonal entries shows that a large top
quark mass can drive one of the top squarks to be quite light.  If the
top squark is light enough (which experiment allows \cite{ehs-3-slite})
 the decay $t \to \tilde{t} \tilde{N}$ becomes possible; that is,
the top quark may be the only quark able to decay to its own
superpartner. 

This idea can be tested in top quark production experiments.  The
simplest test is to see whether the measured top mass and 
cross-section match, since the latter depends on how the top is assumed to
decay.  CDF and DO data indicate that assuming the top decays only
through the standard channel $t \to W b$ gives a good fit to the data
\cite{ehs-3-cdfdo-top}.  If additional decay channels for the top existed, the
production cross-section measured in the $Wb Wb$ final states would be
lower than the standard model prediction.  However, this effect could be
balanced \cite{ehs-3-stops} in supersymmetric models either by production of
states that decay to top quarks $(\tilde{g} \to t \tilde{t})$ or by
production of states whose decays mimic those of top quarks $(\tilde{t}
\to b \tilde{C})$.  Checking these possibilities would require seeking,
specific signatures of the presence of supersymmetric particles; for
instance, while QCD produces mainly top anti-top pairs, gluino pair
production with subsequent decay to top quarks and squarks also produces
top/top and anti-top/anti-top pairs in the ratio [$t \bar{t} : t t :
\bar{t} \bar{t} \approx 2 : 1 : 1$].

\section*{Conclusions}

The top quark's large mass singles it out in several ways.  It may play
a special role in electroweak symmetry breaking (through its effects on
RG running in supersymmetric models or through formation of top quark
condensates).  It has potentially large effects on precision electroweak
observables like $\Delta\rho$ or $R_b$.  In some cases it has a strong
influence on the masses of other particles such as higgs bosons or
superpartners.  Finally, the top quark may be subject to non-standard
interactions that distinguish it from the up and charm quarks.

As a result, the top quark has already made a difference in our attempts
to understand electroweak and flavor physics.  Is the top quark actually
different in any of the ways outlined here ?  Time and experiment
will tell!


\begin{references}
\bibitem{ehs-1-susy} A recent review is S. Dawson, hep-ph/9602229.
\bibitem{ehs-1-stab} G. Anderson, D. Castano, and A. Riotto,
  {\it Phys.Rev.} {\bf D55}, 2950 (1997);
  H. Murayama and M. Peskin, {\it Ann. Rev. Nucl. Part. Sci.} 1996. 
\bibitem{ehs-1-dynam} A recent review is R.S. Chivukula, hep-ph/9701322.
\bibitem{ehs-1-top} Recent reviews of top quark properties and 
  prospects include R. Frey et al., hep-ph/9704243; S. Frixione et. 
  al, hep-ph/9702287.
\bibitem{ehs-2-susyy} H. Baer et al. {\it Phys. Rev.} {\bf D54}, 5866
  (1996) and {\bf D53}, 6241 (1996) and {\bf D52}, 2746 (1995);
  M. Machacek and M. Vaughn, {\it Nucl. Phys.} {\bf B222}, 83 (1983);
  C. Ford et al., {\it Nucl. Phys.} {\bf B395}, 17 (1995); M. Dine,
  A. Nelson, and Y. Shirman, {\it Phys. Rev.} {\bf D51}, 1362 (1995);
  M. Dine. et al., {\it Phys. Rev.} {\bf D53}, 2658 (1996); J. Amundson
  et al., hep-ph/9609374.
\bibitem{ehs-2-evolve} L. Ibanez, {\it Nucl. Phys.} {\bf 218},514
  (1983); {\it Phys. Lett.} {\bf B110}, 215 (1982); J. Ellis,
  D. Nanopoulos, and K. Tamvakis, {\it Phys. Lett} {\bf B121}, 123
  (1983); L. Alvarez-Gaume, J. Polchinski, and M. Wise, {\it
    Nucl. Phys. B221}, 495 (1983); B. Ananthanarayan, G. Lazarides, and
  Q. Shafi, {\it Nucl. Phys.} {\bf D44}, 1613 (1991).
\bibitem{ehs-2-higgsm} J. Ellis, G. Ridolfi and F. Zwirner, {\it
    Phys. Lett.} {\bf B257}, 83 (1991); H.E. Haber and R. Hempfling,
  {\it Phys. Rev. Lett.} {\bf 66}, 1815 (1991).
\bibitem{ehs-2-mr} L.J. Hall and L. Randall, {\it Nucl. Phys.} {\bf B352},
  289 (1991); L. Randall and N. Rius, {\it Phys. Lett.} {\bf B286}, 299
  (1992); N. Rius and E.H. Simmons, {\it Nucl. Phys.} {\bf B416}, 722
  (1994); E.H. Simmons and Y. Su, {\it Phys. Rev.} {\bf D54}, 3580 (1996).
\bibitem{ehs-2-lgh} J.L. Feng, N. Polonsky, and S. Thomas,{\it
    Phys.Lett.} {\bf B370}, 95 (1996). 
\bibitem{ehs-2-dirr} T. Appelquist et al., {\it Phys. Rev. Lett} 
{\bf 53}, 1523 (1984) and {\it
    Phys. Rev.} {\bf D31}, 1676 (1985).
\bibitem{ehs-2-sbs} R.S. Chivukula, S.B. Selipsky and E.H. Simmons, {\it
    Phys. Rev. Lett.} {\bf 69}, 575 (1992).
\bibitem{rbvalue} LEP Electroweak Working Group and SLD Heavy Flavor
  Group (D. Abbaneo et al.), ``A Combination of Preliminary Electroweak
  Constraints on the Standard Model,'' CERN-PPE-96-183, Dec 1996. 
\bibitem{ehs-2-ncetc} R.S. Chivukula, E.H. Simmons, and J. Terning, {\it
    Phys. Lett.} {\bf B331}, 383 (1994); ibid. {\it Phys. Rev.} {\bf
    D53}, 5258 (1996).
\bibitem{ehs-3-wtb} T. Stelzer and S. Willenbrock, {\it Phys. Lett.} 
   {\bf B357}, 125 (1995); E.H. Simmons, {\it Phys. Rev.} {\bf D55},
   5494  (1997).
\bibitem{ehs-3-wtbacc} A.P. Heinson, hep-ex/9605010; A.P. Heinson,
  A.S. Belyaev and E.E. Boos, hep-ph/9612424; M.C. Smith and
  S. Willenbrock,  {\it Phys. Rev.} {\bf D54}, 6696 (1996).
\bibitem{ehs-3-ttcond} V.A. Miransky, M. Tanabashi and K. Yamawaki, {\it
    Phys. Lett.} {\bf B221}, 177 (1989) and {\it Mod. PHys. Lett.} {\bf
    A4}, 1043 (1989); Y. Nambu, EFI-89-08 (1989) unpublished; 
  W.J. Marciano, {\it Phys. Rev. Lett.} {\bf 62}, 2793 (1989);
  W.A. Bardeen, C.T. Hill and M. Lindner, {\it Phys. Rev.} {\bf D41},
    1647 (1990); C.T. Hill, {\it Phys. Lett.} {\bf B266}, 419 (1991).
\bibitem{ehs-3-tc2} C.T. Hill, {\it Phys. Lett.} {\bf B345}, 483 (1995).
\bibitem{ehs-3-njl} Y. Nambu and G. Jona-Lasinio, {\it Phys. Rev.} {\bf
    122}, 345 (1961).
\bibitem{ehs-3-tc2phen}K. Lane and E. Eichten, {\it Phys. Lett.} {\bf
    B352}, 382 (1995); R.S. Chivukula, B.A. Dobrescu, and J. Terning, {\it
    Phys.Lett.} {\bf B353}, 289 (1995); G. Buchalla et al., {\it Phys.Rev.} {\bf D53}, 5185 (1996). 
\bibitem{ehs-3-rscjtbd} R.S. Chivukula and J. Terning, {\it
    Phys.Lett.} {\bf B385}, 209 (1996).
\bibitem{ehs-3-slite} D0 Collaboration (S. Abachi et al.). {\it
    Phys.Rev.Lett.} {\bf 76}, 2222 (1996). 
\bibitem{ehs-3-cdfdo-top} M. Paulini (for the CDF and D0 Collaborations),
  hep-ex/9701019. 
\bibitem{ehs-3-stops} G.L. Kane and S. Mrenna, {\it Phys.Rev.Lett.} {\bf
    77},  3502 (1996).  

\end{references}
\end{document}